\documentclass[apj]{emulateapj}
\newcommand{\atom}[2]{\ensuremath{{}^{#2}\mathrm{#1}}}
\newcommand{\nehe}{\ensuremath{\mathrm{Ne/He}}}
\newcommand{\nep}{\ensuremath{\mathrm{Ne^+}}}
\newcommand{\nepp}{\ensuremath{\mathrm{Ne^{++}}}}
\newcommand{\nephe}{\ensuremath{\mathrm{Ne^+/He}}}
\newcommand{\nes}{\ensuremath{\mathrm{Ne/S}}}
\newcommand{\she}{\ensuremath{\mathrm{S/He}}}
\newcommand{\sppp}{\ensuremath{\mathrm{S^{3+}}}}
\newcommand{\sppphe}{\ensuremath{\mathrm{\sppp/He}}}
\newcommand{\nepsppp}{\ensuremath{\mathrm{Ne^+/S^{3+}}}}
\newcommand{\msun}{\ensuremath{\mathrm{M}_\odot}}
\newcommand{\heI}{He\,\textsc{i}}
\newcommand{\heII}{He\,\textsc{ii}}
\newcommand{\neII}{[Ne\,\textsc{ii}]}
\newcommand{\neIII}{[Ne\,\textsc{iii}]}

\newcommand{\sIV}{[S\,\textsc{iv}]}

\newcommand{\lene}{\ensuremath{\log(\epsilon_{\mathrm{Ne}})}}
\newcommand{\um}{\,\micron}
\newcommand{\score}{\textsc{Score}}

\defcitealias{2001AJ....121.2115S}{SH01}

\lefthead{Smith \& Houck}  \righthead{Neon Abundance of Galactic WR Stars}

\begin{document}

\title{The Neon Abundance of Galactic Wolf--Rayet Stars}

\author{John-David T. Smith}
\affil{Steward Observatory, The University of Arizona, Tucson, AZ 85721}
\email{jdsmith@as.arizona.edu}

\and

\author{James R. Houck}
\affil{Center for Radiophysics and Space Research, Cornell University, 
  Ithaca, NY 14853}

\begin{abstract}
  The fast, dense winds which characterize Wolf-Rayet (WR) stars obscure
  their underlying cores, and complicate the verification of evolving
  core and nucleosynthesis models.  Core evolution can be probed by
  measuring abundances of wind-borne nuclear processed elements,
  partially overcoming this limitation.  Using ground-based mid-infrared
  spectroscopy and the 12.81$\mu m$ \neII\ emission line measured in
  four Galactic WR stars, we estimate neon abundances and compare to
  long-standing predictions from evolved-core models. For the WC star
  WR\,121, this abundance is found to be $\gtrsim 11\times$ the cosmic
  value, in good agreement with predictions.  For the three less-evolved
  WN stars, little neon enhancement above cosmic values is measured, as
  expected.  We discuss the impact of clumping in WR winds on this
  measurement, and the promise of using metal abundance ratios to
  eliminate sensitivity to wind density and ionization structure.
\end{abstract}

\keywords{infrared radiation --- stars: Wolf--Rayet --- techniques:
spectroscopic}

\section{INTRODUCTION}

Wolf--Rayet (WR) stars are evolved massive stars characterized by high
mass loss rates ($10^{-5}\le\dot{M}\le10^{-4}\:\msun\:\mathrm{yr}^{-1};
\dot{M}_{WR}\sim10^{10}\:\dot{M}_\odot$), driven in fast ($v_\infty \sim
\mathrm{few} \times 10^3\,\mathrm{km/s}$) stellar winds.  The dense WR
winds obscure their underlying cores and the region from which the
outflowing wind material is initially accelerated.  The difficulty in
quantifying bulk parameters of WR stars which arises due to this
obscuration is typified by the uncertainty regarding the appropriate
photospheric temperature to assign them.  When the most commonly used
definition of the photosphere is adopted (the location of optical depth
$\tau\sim2/3$), a temperature degeneracy arises in the models of WR
atmospheres --- for all WR subtypes, the effective photospheric
temperature derived is $T_{2/3}\sim30,000\,\mathrm{K}$, for reasons
unknown \citep*{Schmutz1992}.  More effective for predicting luminosity
and emergent flux distribution is the inferred \emph{core} temperature,
but this assignment depends critically on the \emph{ad hoc} choice of a
velocity structure of the wind, the commonly assumed form of which
\citet{Schmutz1997} showed to be largely invalid in the single WR star
(HD 50896 -- WR\,6) for which the optically thin, supersonic portion of
the wind has been modeled hydrodynamically.  Testing models of WR core
evolution, and the advanced nuclear reactions which occur there, is
complicated by this disconnect between the observable bulk properties of
the wind, and the wind-driving core buried beneath it.

A powerful technique for probing WR core evolution which sidesteps these
difficulties is available in the measured abundances of wind-borne
nuclear processed elements.  Neon in particular undergoes a remarkable
abundance change during the later stages of a WR star's lifetime.  By
the end of the more evolved WC phase (characterized by wind material
dominated by $\alpha$-burning by-products, carbon in particular),
\atom{Ne}{22} becomes the fourth most abundant element, after
\atom{He}{4}, \atom{C}{12}, and \atom{O}{16}.  The reactions of interest
contributing to the creation of neon during He-burning in massive stars
are \citep{Maeder1983}:

\vspace{1em}
\noindent\mbox{
  \atom{N}{14}$(\alpha,\gamma)\Rightarrow$%
  \atom{F}{18}$(\beta,\nu)\Rightarrow$%
  \atom{O}{18}$(\alpha,\gamma)\Rightarrow$%
  \atom{Ne}{22}%
  \parbox{1.5in}{
    $(\alpha,n)\Rightarrow$%
    \atom{Mg}{25}\\
    $(\alpha,\gamma)\Rightarrow$%
    \atom{Mg}{26}
  }
}

\vspace{1em}

\noindent\mbox{
  \atom{He}{4}$(\alpha,\gamma)\Rightarrow$%
  \atom{Be}{8}$(\alpha,\gamma)\Rightarrow$%
  \atom{C}{12}$(\alpha,\gamma)\Rightarrow$%
  \atom{O}{16}$(\alpha,\gamma)\Rightarrow$%
  \atom{Ne}{20}
}
\vspace{1em}

\noindent Essentially all of the \atom{N}{14} produced via the CN
branch of the CNO cycle which dominates the earlier WN evolutionary
phase is converted to \atom{Ne}{22}.  The further conversion of neon to
\atom{Mg}{25} and \atom{Mg}{26} is inefficient except at the highest
temperatures (achieved only in stars with initial masses
$\gtrsim100\textrm{M}_\odot$), and the production of \atom{Ne}{20} via
\atom{O}{16} is also negligible except in the most advanced WC and WO
stages of the highest mass stars.  The two main consequences of these
critical neon production chains are a strong increase in the overall
abundance of neon by a factor of $\sim$200 over the course of the WR
phase, and a rise in the isotopic abundance ratio
\atom{Ne}{22}/\atom{Ne}{20} from $\sim$0.1 to $\sim$35.  Both of these
changes to the neon abundance occur quite rapidly (in the course of
several thousand years) at the onset of the WC phase.  Late-stage
depletion of \atom{Ne}{22} by conversion to magnesium is only 30\%
during the final $10^6$ yr of the most massive stars' lives, with the
overwhelming majority of WRs maintaining their full neon excess, thanks
to the combined effects of the interior mixing and mass-loss which bring
material to the surface.  Assuming normal p-p nuclear processing of
hydrogen entirely to helium, the cosmic abundance of neon by number is
$\nehe=3.7\times 10^{-4}$ (see \S\,\ref{sec:cosm-neon-abund}).  The neon
abundance predicted by WR evolutionary models is $\nehe=6.6\times
10^{-3}$, or over 17 times the cosmic value.  This result, first
described by \citet{Maeder1983}, has remained valid despite recent model
updates to accomodate rotational mixing and the turbulent diffusion of
core material into the wind it drives \citep{Maeder2000}.

We will give background on the detection of neon in WR winds in
\S\,\ref{sec:background}, describe the observations in
\S\,\ref{sec:observations}, outline the abundance calculation and inputs
in \S\S\,\ref{sec:abund-calc}--\ref{sec:targets--inputs}, and present
the results in \S\,\ref{sec:results}. 

\section{BACKGROUND}
\label{sec:background}

Despite lingering uncertainties concerning the true structure of WR
winds, broad agreement between the surface abundances predicted by core
evolution models was obtained early on for almost all abundant elements,
both for WN \citep*[e.g.][]{Crowther1995} and WC
\citep[e.g.][]{Maeder1994a} stars.  However, a long standing discrepancy
concerning the model-sensitive neon abundance predictions centered on
$\gamma^2$ Velorum (WC8), the nearby, optically-brightest WR star.
While the \citeauthor{Maeder1983} abundances were well-confirmed for all
other elements, the measured neon abundances of $\gamma^2$ Vel remained
quite low, near or just above the cosmic value.

In one of the earliest mid-infrared (MIR) measurements of a WR star,
\citet*{Aitken1982}\ calculated Ne$^{+}$ and S$^{3+}$ abundances in
$\gamma^{2}$ Vel from ground-based spectra.  \citet{vanderHucht1985}\ 
found a similarly elevated neon abundance for $\gamma^{2}$ Vel using
IRAS LWS spectra, seemingly confirming its predicted over-abundance with
respect to cosmic levels.  Both of these results, however, were later
significantly corrected by \citet{Barlow1988}, who discovered flaws in
the calculations and in the atomic inputs, and found a revised neon
abundance again quite close to cosmic values.  \citet{Dessart2000}\ 
adjusted $\gamma^{2}$ Velorum's neon abundance yet again, using ISO data
and an improved distance value along with more modern,
clumping-corrected mass loss rates to derive a value coincidentally
quite close to the original determination of \citeauthor{Aitken1982},
and in good agreement with theory.  \citet{Willis1997} found an elevated
neon abundance in the ISO SWS spectrum of WR\,146 (WC5), and in total,
four ISO WC stars \citep{Willis1997,Dessart2000} and one WN star
\citep{Morris2000} have yielded neon abundances.  \citet{Morris2004}
used early Spitzer IRS spectra to measure the neon abundance of WN4 star
WR6, and found values consistent with no enhancement.

\begin{deluxetable*}{llllr@{\,$\pm$\,}lll}
\tablecaption{Program WR Stars with Observed Neon\label{tab:targets}}
\tablecolumns{8}
\tablehead{ 
\colhead{Object} & 
\colhead{Name} & 
\colhead{Spectral} &
\colhead{D} & 
\multicolumn{2}{c}{$A_{v}$} &
\colhead{$v$} &
\colhead{$f_{10\mu m}$}\\
\colhead{} &
\colhead{} &
\colhead{Type} &
\colhead{(kpc)} &
\multicolumn{2}{c}{(mag)} &
\colhead{(mag)} & 
\colhead{(Jy)}}
\startdata
WR\,105 &Ve2-47   & WN9h & 1.58 & 8.63 & .03              & 12.92 & 0.68\\
WR\,116 &ST\,1    & WN8h & 2.48 & \multicolumn{2}{c}{6.89}& 13.38 & 0.44\\
WR\,121 &AS\,320  & WC9d & 1.83 & 5.72 & .32              & 12.41 & 1.51\\
WR\,124 &209\,BAC & WN8h & 3.36 & \multicolumn{2}{c}{4.43}& 11.58 & 0.14
\enddata

\tablecomments{All spectral types, magnitudes, and extinction
  coefficients are from \citet{vanderHucht2001}.  Extinctions with
  quoted errors are for stars with known cluster associations, and are
  considerably more accurate.}
\end{deluxetable*}

\section{OBSERVATIONS}
\label{sec:observations} 

\begin{figure}
\plotone{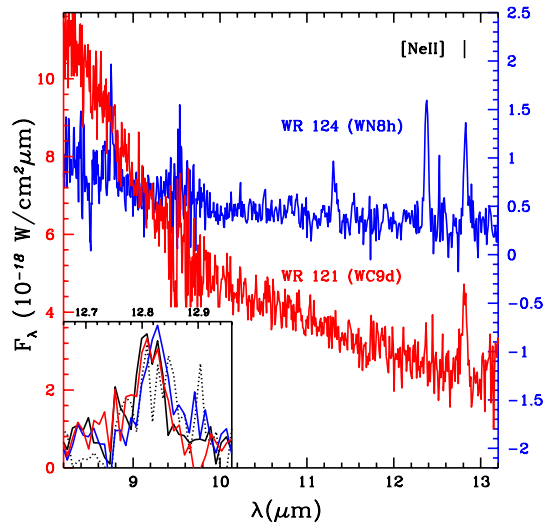}
\caption{\score\ spectra of two of the program stars.  The \neII\
  12.81\um\ line is inset for all four stars (solid: WR105, dotted:
  WR116, see \citetalias{2001AJ....121.2115S} for full spectra).  The
  dusty WC9d has a steep slope typical of $\sim$1000K heated dust
  emission, with a slight break in the slope at 9.5\um, characteristic
  of moderate, broad silicate absorption.  The noise excess in both
  spectra at 9.7\um\ is due to telluric O$_3$.
  \label{fig:wrspec}}
\end{figure}

\citet{2001AJ....121.2115S} (hereafter \citetalias{2001AJ....121.2115S})
present a flux-calibrated 8--13\um\ spectral survey of a large sample of
northern Galactic WR stars representing all sub-types.  Among the
sample, four stars exhibited broad, non-nebular \neII\,12.81\um\ line
emission.  These are listed in Table~\ref{tab:targets}, along with
spectral types, photometric or cluster distances and reddening, visual
magnitudes and mid-infrared fluxes.  The spectra were obtained with
\score\ \citep{1998SPIE.3354..798S,1998PASP..110.1479V}, a prototype
instrument for the short wavelength, high resolution module of Spitzer's
IRS spectrograph \citep{Houck2004}, operated at the Palomar
5m\footnote{Observations at the Palomar Observatory were made as part of
  a collaborative agreement between the California Institute of
  Technology, the Jet Propulsion Laboratory and Cornell University.}.
Standard beam-switched 5Hz chopping and nodding were used to remove the
sky signal.  The chop amplitude was chosen to be small enough so that
the object fell within \score's 12\arcsec\ diameter slit-viewer field
when the slit was on adjacent sky.  Two equal amplitude source images in
the slit-viewer's 11.3\um\ silicate filter were thus obtained
simultaneously with the spectra, and used to correct the absolute flux
calibration for the changing slit throughput function, which is affected
by seeing, pointing accuracy, and object acquisition.  This calibration,
and a general description of the spectral reduction process, are
described in more detail by \citetalias{2001AJ....121.2115S}.  Initial
flux-calibration was performed using same-airmass observations of
\citet{Cohen1995} infrared standard stars.  Line strengths were computed
using a polynomial fit to the nearby continuum.  Fig.~\ref{fig:wrspec}
shows the \score\ MIR spectra of one of the WN and the single WC program
stars.  The WN star exhibits lines of helium and neon, while the WC star
exhibits only \neII.

One distinct difference between the \score\ data used here and ISO SWS
spectra used to compute previous neon abundances is worth mentioning.
The entirety of ISO's SWS spectral beam ($\sim20\arcsec\times35\arcsec$)
is mapped onto the spectrograph's large detector elements and is
included in the recorded spectra.  For this reason, contamination by
\emph{nebular} emission lines originating in the often bright, extended
nebulae surrounding WR stars is often a concern.  \score's small
$1\arcsec\times2\arcsec$ slit excludes most of this nebular emission,
leading in some cases to significant differences between spectra of the
same non-variable WR star observed with both ISO and \score\ (cf. WR146
in \citet{Willis1997} vs.  \citetalias{2001AJ....121.2115S}, as
described in \S4.2 of the latter).  While we know of no cases in which
neon or other abundances computed using ISO data were affected by this
type of contamination, the \score\ spectra of fainter WR stars used here
should be relatively less affected by nebular emission.

\section{CALCULATING NEON ABUNDANCE}
\label{sec:abund-calc}

\begin{figure}
\plotone{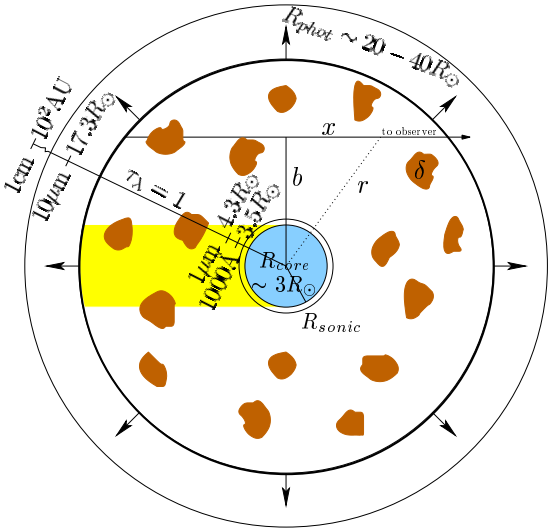}
 \caption{Wind model geometry.  Observer is to the
   right, along the line of sight coordinate $x$.  The clumps of fill
   factor $\delta$ are indicated schematically.  The shading to the left
   of the core along the line of sight indicates the region from which
   no continuum emission is observed.  Also indicated are representative
   physical radii for the core, sonic point, and the photosphere
   \citep[from][]{Heger1996}, as well as the location of the continuum
   optical depth $\tau_\lambda=1$ at a variety of wavelengths, from the
   WN5 wind model of \citet{Hillier1987}.
    \label{fig:windmodel}}
\end{figure}


\subsection{Mass Loss \& Clumping}
\label{sec:mass-loss-clump}

For a uniform, spherical, but clumped wind with terminal velocity
$v_{\infty}$ and constant volume filling fraction $\delta$
({$0\le\delta\le1$), as depicted in Fig.~\ref{fig:windmodel}, the mass loss
  rate can be written:
\begin{equation}
  \label{eq:1}
  \dot{M}=4\pi r^2 \delta\cdot\!n\mu m_H v_{\infty}
\end{equation}
where $\mu$ is the mean atomic mass per ion, and $n$ is the number
density of the ionized gas.  Defining the standard \emph{mass loss
  coefficient} $\mathcal{A}\equiv \frac{\dot{M}}{4\pi\mu
  m_Hv_{\infty}}$, the density can be expressed as:
\begin{equation}
  \label{eq:2}
  n=\frac{\mathcal{A}}{\delta r^2}
\end{equation}%

The dominant radiative output of the wind at mid-infrared and longer
wavelengths is free-free emission \citep{Wright1975}.  The free-free
optical depth along a particular line of sight through the clumped wind
to the observer is

\begin{equation}
  \label{eq:3}
  \tau(b)=\int \delta \kappa_{ff}dx=\delta\mathcal{K}(\nu,T)\int n_en\, dx=\frac{\pi\mathcal{K}(\nu,T)\mathcal{A}^2\gamma_e}{2\delta b^3}
\end{equation}
where the reduced free-free opacity
$\mathcal{K}(\nu,T)=\kappa_{ff}/nn_e$, $\gamma_e$ is the number of
electrons per ion, and we have made use of the $1/r^2$ density profile
of Eq.~\ref{eq:2}.  Assuming a constant, thermal source function, and
integrating over cylinders of constant impact parameter $b$ (and thus
constant free-free optical depth), we recover \citeauthor{Wright1975}'s
infrared/radio free-free flux expression, modified to include the
effects of clumping via the fill factor $\delta$:

\begin{equation}
  \label{eq:4}
  S_\nu=
  \left(\frac{\dot{M}Z}{\sqrt{\delta}\mu v_\infty}\right)^{4/3}
  \frac{\left(\gamma_eg_\nu \nu\right)^{2/3}}{D^2}
  \left(\frac{\pi e^6}{6\sqrt{3}c^4m_H^2m_e^{3/2}}\right)^{2/3}
  \Gamma(\onethird)
\end{equation}
where $D$ is the distance to the star, $g_\nu$ is the frequency
dependent free-free Gaunt factor, and $Z$ is the rms average charge per
ion.  Most WR mass loss rate estimates are derived from radio
measurements of the free-free emission using Eq.~\ref{eq:4}.  Given the
same assumptions of atomic parameters of the wind ($Z$, $\mu$,
$\gamma_e$), it is apparent that, in the absence of information about
the clumping fill factor $\delta$, the \emph{clumping-scaled mass loss
  rate}, $\dot{M}_{scl}\equiv\dot{M}/\sqrt{\delta}$, is derived.

\subsection{Two level emission}
\label{sec:two-level-emission}

For fine structure lines arising from ions with ground states consisting
of only two energy levels, the fractional abundance of the ion by
number, \mbox{$\gamma_i\equiv n_i/\sum_j n_j$}, can be calculated
straightforwardly from the observed line flux by neglecting all other
transitions.  Following \citet{Barlow1988}, the flux due to a given line
transition can be written:

\begin{equation}
  \label{eq:5}
  4\pi D^2\mathcal{F}_{ul}=\int\limits_0^\infty n_uA_{ul}h\nu_{ul}
  4\pi r^2 dr
\end{equation}
where $A_{ul}$ is the Einstein emission coefficient for the line in
question, and $n_u$ is the density of the ions populating the upper
level of the transition.

The upper level density and electron density follow from the definition
of $\gamma_i$:
\begin{equation}
  \label{eq:6}
  n_u=\frac{f_u\gamma_i\mathcal{A}}{\delta r^2} \:;\: 
  n_e=\frac{\gamma_e\mathcal{A}}{\delta r^2}
\end{equation}
where $f_u$ is the fraction of that ionic species present in the upper
level.  Plugging into Eq.~\ref{eq:5} for $n_u$, we find the line flux to
be:
\begin{equation}
  \label{eq:7}
  \mathcal{F}_{ul}=\frac{\gamma_i\mathcal{A}A_{ul}h\nu_{ul}}{D^2}
  \int\limits_0^\infty f_u(n_e(r),T)\:dr
\end{equation}
To derive the upper-level fraction $f_u(n_e(r),T)$, we begin with the
total collisional de-excitation rate per unit volume is
\citep{Osterbrock1974}:
\begin{equation}
  \label{eq:8}
  n_en_uq_{ul}=\sqrt{\frac{2\pi}{kT}}\frac{\hbar^2}{m^{\frac{3}{2}}}
  \frac{\Omega_{ul}(T)n_en_u}{\omega_u}
\end{equation}
where $\Omega_{ul}(T)$ is the collision strength, and $\omega_u$ is the
statistical weight of the upper level.  The collisional excitation rate
can be obtained from the de-excitation rate using detailed balance:
\begin{equation}
  \label{eq:9}
   n_en_lq_{lu}=\sqrt{\frac{2\pi}{kT}}\frac{\hbar^2}{m^{\frac{3}{2}}}
  \frac{\Omega_{lu}(T)n_en_l}{\omega_l}e^{\frac{-h\nu_{ul}}{kT}}
\end{equation}
In the simple two level ion, statistical equilibrium between the levels
can be written succinctly:
\begin{equation}
  \label{eq:10}
  n_en_lq_{lu}=n_en_uq_{ul}+A_{ul}n_u
\end{equation}
from which the upper level population can be found simply, using
$n_i=n_u+n_l$, as
\begin{equation}
  \label{eq:11}
  f_u=\left(1+\frac{1+\frac{n_c}{n_e}}
    {\frac{q_{lu}}{q_{ul}}}\right)^{-1}
\end{equation}
where $n_c\equiv\frac{A_{ul}}{q_{ul}}$ is the critical density in
the two level approximation.  Plugging in for $n_e$ from
Eq.~\ref{eq:6}, and the ratio of the collisional rates from
Eqs.~\ref{eq:8}~--~\ref{eq:9}, we find
\begin{equation}
  \label{eq:12}
  \mathcal{F}_{ul}=\frac{\gamma_i\mathcal{A}A_{ul}h\nu_{ul}}{D^2}
  \int\limits_0^\infty 
  \left(1+\left[1+\frac{n_c\delta r^2}{\gamma_e\mathcal{A}}\right]
    \frac{\omega_l}{\omega_u}e^{\frac{h\nu}{kT}}
    \right)^{-1}dr
\end{equation}

\noindent from which we can solve for the fractional abundance by number
of the ion contributing to the line emission:

\begin{equation}
  \label{eq:13}
  \gamma_i=\frac{2\mathcal{F}_{ul}D^2}
  {\pi \mathcal{A}^{\frac{3}{2}}A_{ul}h\nu}\frac{\omega_l}{\omega_u}
  e^{\frac{h\nu}{kT}}\sqrt{\frac{\delta n_c}{\gamma_e}
  \left[\frac{\omega_u}{\omega_l}e^{-\frac{h\nu}{kT}}+1\right]}
\end{equation}

Notice that $\gamma_i\propto\sqrt{\delta}/\mathcal{A}^{\frac{3}{2}}$.
In the absence of information about the clumping fill factor $\delta$,
the scaled mass loss rate is derived from radio continuum measurements
(see \S\,\ref{sec:mass-loss-clump}).  If the scaled mass loss rate is
known then $\mathcal{A}\propto\dot{M}=\dot{M}_{scl}\sqrt\delta$, and the
weak scaling of fractional abundance with clumping factor becomes
$\gamma_i\propto\delta^{-\frac{1}{4}}$.

Note that we differ from \citet{Dessart2000} and \citet{Morris2000}, who
perform a numerical integration over the upper level ionic fraction
$f_u$, after transforming from radial to density coordinates to mitigate
finite step size inaccuracies near the origin.  While they find
discrepancies of order 20\% between their analytical integral analogous
to Eq.~\ref{eq:12} and their numerical integration, we find no reason
that such differences should occur for this two-level transition (for
which the upper level fraction integral is exact) and presume it must
have arisen from different atomic data inputs in the statistical
equilibrium code used, or the finite numerical resolution of the
integration.

Typically, WR wind abundances are formulated with respect to helium,
almost always the most abundant element in all but the least evolved,
late-type WN winds.  Given knowledge of other elemental abundances, this
can be calculated according to:

\begin{equation}
  \label{eq:14}
  \mathrm{\frac{X_i}{He}}=\gamma_i\mathrm{
      \left(1+\frac{H}{He}+\frac{C}{He}+
        \frac{N}{He}+\frac{O}{He}+\ldots\right)}
\end{equation}
where $\mathrm{X}_i$ represents the ion $i$, and the standard shorthand
$\mathrm{X/Y}\equiv \mathrm{N_{X}/N_{Y}}$ has been used.  Only the
most important additional elements beyond helium are shown, and in many
cases, the abundances of most will be so low that only one other element
besides helium need be considered (e.g. carbon for WC stars, nitrogen
and/or hydrogen for WN stars).

\section{INPUTS}
\label{sec:targets--inputs}
\begin{deluxetable*}{lccccccc}
\tablecolumns{8}
\tablewidth{0pc} 

\tablecaption{Atomic Data for Neon and Sulfur Lines.\label{tab:atomic}}

\tablehead{ 
\colhead{Ion} & 
\colhead{Transition} & 
\colhead{Wavelength} &
\colhead{$\omega_l$}&
\colhead{$\omega_u$}&
\colhead{$\Omega_{ul}$}&
\colhead{$A_{ul}$} & 
\colhead{$n_{c}(8000\mathrm{K})$} \\
\colhead{}&
\colhead{}&
\colhead{($\mu$m)}&
\colhead{}&
\colhead{}&
\colhead{}&
\colhead{($10^{-3}\,\mathrm{s}^{-1}$)}&
\colhead{($cm^{-3}$)}}

\startdata
\sIV&$^2\!P^\circ_{3/2}\rightarrow{}^2\!P^\circ_{1/2}$&10.5105&2&4&8.47&7.70&3.77$\times10^4$\\
\neII&$^2\!P^\circ_{1/2}\rightarrow{}^2\!P^\circ_{3/2}$&12.8136&4&2&0.28&8.59&6.36$\times10^5$
\enddata 
\tablecomments{$\Omega_{ul}$ computed at $T_e=8000\mathrm{K}$.
  \citet{Dessart2000}.}
\end{deluxetable*}

The fractional ionic abundance can be calculated directly from
Eq.~\ref{eq:13}, given estimates of the distance, atomic wind
parameters, and reddening.  Table~\ref{tab:atomic} lists the atomic data
used for the two fine-structure lines considered here.  The measured
neon and sulfur line fluxes used here differ slightly from those
presented in \citetalias{2001AJ....121.2115S}, though the reduced
spectra are unchanged.  This is a result of better estimation of the
continuum.

\subsection{Chemical Composition}
\label{sec:chemical-composition}

The adopted chemical composition of the winds serves only to normalize
the computed abundance with respect to helium (Eq.~\ref{eq:14}), but
does not change the results otherwise.  All abundances mentioned are by
number.

The three WN stars with neon present are all late types --- the only WR
type with any significant hydrogen remaining.  The abundances assumed
follow \citet*{Nugis1998}, with \mbox{N/He=0.005} for all late WN's.  The
hydrogen content (\mbox{H/He}) of WR\,105 (2.3) and WR\,124 (1.9) were
available from \citet{Nugis2000}, computed using optical \heI, \heII\ 
and H\,\textsc{i} line measurements.  The latter value is also adopted
for the WN8h star WR\,116.

For the carbon star WR\,121, the weighted mean in \citeauthor{Nugis2000}
for subtypes WC8-9 of carbon abundance (\mbox{C/He=0.18}), as well as
oxygen abundance (\mbox{O/C=0.2}) was used.  The WC9's hydrogen
abundance was assumed to be zero.

\subsection{Mass Loss and Terminal Velocity}
\label{sec:mass-loss-terminal}

The mass loss rate enters Eq.~\ref{eq:13} through the mass loss
coefficient as $\mathcal{A}^{3/2}$, and hence significantly affects the
values obtained for the abundance.  Unfortunately, rates derived for
individual stars often differ substantially, depending on the input
assumptions, and details of the measurement.  Along with imprecise
distance estimates, poorly-constrained mass loss rates introduce the
largest uncertainties in the computed abundances.

Eq.~\ref{eq:4} can be used to determine the mass loss rate (or at least
the scaled rate --- $\dot{M}/\sqrt{\delta}$) from the measured radio
flux, given independent estimates of the distance, terminal expansion
velocity, and atomic parameters of the wind
\citep[e.g.][]{Leitherer1995,Leitherer1997}.  The radio emitting regime
is quite far out in the wind, such that the assumption that the terminal
velocity has been obtained is always valid.  The infrared free-free
continuum, however, may not be optically thick (recall that
$\kappa_{ff}\propto\lambda^3$ in the long wavelength limit).
Contributions to the 10\um\ continuum from the underlying stellar
photosphere are typically only 10\% of the free-free wind continuum
\citep{Barlow1981}, and the cross-over from photosphere to
wind-dominated emission occurs near 1--5\um; however, since the
free-free opacity is lower, the infrared emission originates deeper in
the wind, sampling outflow which may not yet have reached terminal
velocity, such that the assumption leading to Eq.~\ref{eq:2} and a
$1/r^2$ density distribution is violated, and the spectral index
steepens compared to the constant velocity limit.\footnote{A simple
  understanding of why the spectral index must steepen in an
  accelerating wind is had by noting that at different frequencies,
  $\tau_{ff}\sim 1$ occurs at different depths, since
  $\frac{d\tau}{ds}=\kappa_{ff}$ is quadratic in the density.  If the
  rate of change of density with radius is steeper (as in a region of
  acceleration), adjacent frequencies will originate further apart in
  the flow, exaggerating the difference.}  In the absence of a valid
velocity law, one possible technique to measure the mass loss rate is to
scale the infrared flux with an empirically determined spectral index
$\alpha$ ($f_\nu\propto\nu^\alpha$), as \citet{Barlow1981} did for
$\gamma$ Velorum and HD 192163 to estimate a scaling with $\alpha=0.76$.
More recently, \citet{Crowther1995} found a value in close agreement,
$\alpha=0.74$, based on a tailored analysis of several WN stars.

Scaling our MIR fluxes with this method and spectral index
$\alpha=0.74$, mass loss rates for the three program WN stars (i.e.
those without clear evidence of dust emission) were calculated.  Each
spectrum was dereddened using a composite extinction curve of Cohen
(priv. communication), which, longward of $4.7\um$, joins smoothly the
curves used by \citet{Cohen1993}.  The small inferred MIR reddening
corrections are relatively uncertain, but the largest uncertainty is for
the dusty types not considered in this calculation.

A 0.2\um\ wide region centered on 12\um\ was averaged to arrive at the
\score\ flux density.  IRAS Point Source Catalog 12\um\ flux densities
\citep{Cohen1995b} were found to be 2--5$\times$ higher, likely due to
the very large IRAS beamsize.  The extrapolation of the de-reddened
12\um\ fluxes was performed to 4.80\,GHz (6.25\,cm) for comparability
with \citet{Leitherer1997}.  Atomic parameters of the wind were adopted
from \citeauthor{Leitherer1997}, using the usual relation for mean ionic
mass
\begin{equation}
  \label{eq:58}
  \mu=\frac{\sum \gamma_im_i}{\sum \gamma_i}
\end{equation}
where $m_i$ is the atomic mass of the ion.  Only H, He, and C are
included, since O never reaches abundances which would influence $\mu$
significantly.  The mean molecular weights per ion
\citeauthor{Leitherer1997} compute are consistent enough within subtypes
that we adopt $\mu=2.0$ for WN types later than WN6.

The mean number of electrons per ion, $\gamma_e$, and the RMS ionic
charge, $Z$, were computed assuming He$^+$ is the most prevalent ion in
the radio emitting region.  These quantities are very insensitive to any
other assumption, since singly ionized helium and hydrogen contribute to
both equivalently.  Carbon is often assumed to exist as C$^{++}$ in the
radio regime, but it does not significantly impact the calculation of
either $\gamma_e$ or $Z$, each of which typically only ranges from
1.0--1.2.  Given the magnitude of other uncertainties in the
calculations, we adopt $Z=\gamma_e=1.1$.  The Gaunt factor, which enters
Eq.~\ref{eq:4} through the free-free opacity, depends logarithmically on
temperature and ionic charge at 4.90\,GHz .  We again follow
\citeauthor{Leitherer1997} in adopting $g_{4.8}=5.0$, which they
computed for $T_e=10,000\,\mathrm{K}$.

\begin{deluxetable}{lr@{.}lc}
\tablecolumns{4} 
\tablewidth{0pc} 
\tablecaption{Example Mass-Loss Rates from Various
Methods.\label{tab:mass-loss-examp}}
\tablehead{ 
\colhead{Method}&
\multicolumn{2}{c}{$\log(\dot{M})\:(\mathrm{M_{\odot}/yr})$}&
\colhead{Reference}}

\startdata  
\cutinhead{WR\,105 (WN9h)}
Radio                           & $<$-4&41 & (1)\\
Radio+Clumping                  & -4&55    & (2)\\
Extrapolated IR                 & -4&95    & (4)\\
Optical+1\um\ Line Analysis    & -4&1     & (5)\\
Optical+UV Line Analysis        & -4&2     & (6)\\
\cutinhead{WR\,124 (WN8h)}
Optical Line+Clumping           &  -4&61   & (2)\\
Extrapolated IR                 &  -4&7    & (3)\\
Extrapolated IR                 &  -4&95   & (4)\\
Optical+UV Line Analysis        &  -3&8    & (6)
\enddata

\tablerefs{(1) \citet{Leitherer1997}. (2) \citet{Nugis2000}. (3)
  \citet{Barlow1981}.  (4) this work.  (5)
  \citet*{Schmutz1989}.  (6) \citet{Hamann1998b}}
\end{deluxetable}

For two of the sources, mass loss rates were available from the
clumping-corrected emission line-fitting results of \citet{Nugis2000}:
$\log(\dot{M})=-4.55$ (WR\,105) and $\log(\dot{M})=-4.61$ (WR\,124).
Illustrating the uncertainty in the rates,
Table~\ref{tab:mass-loss-examp} lists different estimates based on
different techniques for two of the WN program stars.  Rates based on
UV, optical, IR, and radio data, with and without clumping accounted
for, are listed for two of our program stars.  It should be pointed out
that only the \emph{Radio} and \emph{Extrapolated IR} methods do not
require detailed modeling with the implicit \emph{ad hoc} assumption for
the form of the velocity field, although the IR rates do depend
sensitively on the measured spectral slope for extrapolation.  The
variance among the different estimates is quite large, and, in the case
of WR\,124, the smallest and largest rates found differ by a factor of
$\sim$14.

WR\,105 is a confirmed non-thermal emitter \citep{Chapman1999}, with a
spectral index $\alpha=-0.3$ over the 3--6\,cm radio band.  Since the
infrared flux will likely not be modified by the thermal X-ray and
non-thermal radio emission arising in the binary wind-wind collisional
shocks thought to underly non-thermal WR sources \citep{Eichler1993}, we
expect this star's 12\um\ extrapolated mass loss rate
($\log(\dot{M})=-4.95$) to be more accurate than radio-based results,
and adopt it for abundance calculations.

Where available, we have preferred the radio and extrapolated infrared
rates over others.  In one case (WR\,116) no radio or clumping corrected
results were available, and we therefore adopted $\log(\dot{M})=-4.6$,
based on observations of other WN8 stars.

\citet{Leitherer1997} put an upper limit on the mass loss rate of
WR\,121 from Australia Telescope Compact Array radio non-detection of
$\log(\dot{M})<-4.5$, and \citet{Bieging1982} find a similar limit
($<$-4.55) using the Very Large Array.  \citet{Abbott1986}, however, had
previously placed a firmer limit of $\log(\dot{M})<-4.8$.  We adopt
$\log(\dot{M})=-4.9$, based on values derived from radio detections of
other WC9 stars, and in line with the lower mass-loss rate WC9's exhibit
compared to earlier WCs \citep{Leitherer1997}.  The rate inferred from
scaling WR\,121's MIR flux is not valid, having been unduly influenced
by the excess heated dust emission evident in WR\,121's spectrum.

All terminal velocities are from \citet{vanderHucht2001}, except for
WR\,105, for which we adopt the value of \citet{Eenens1994},
$v_\infty=1200\,\mathrm{km/s}$, which closely matches our measurements
of the \sIV\ 10.5\um\ FWHM in our spectrum, but deviates from most the
recently cataloged value of $700\,\mathrm{km/s}$.

\subsection{Temperature and Atomic Parameters}
\label{sec:temp-atom-param}

Despite the very high inferred effective temperatures ($>$60\,kK) of the
underlying star, WR winds are quite efficiently cooled by line radiation
to $T_e\lesssim10\mathrm{kK}$ in the radio emitting regime
\citep{Hillier1989}.  Though the exact temperature in the line-emitting
region is uncertain, the temperature dependence of the derived ionic
abundance is extremely weak at these long wavelengths
($h\nu/kT_e\sim.1\Rightarrow\gamma_i\propto T_e^{0.07}$); we therefore
assume $T_e=8000\mathrm{K}$ for all four sources.

The mean ionic masses ($\mu$) and mean number of electrons per ion
($\gamma_e$) were taken from \citet{Leitherer1997} in the case of WR\,105
and WR\,121.  For WR\,124, $\mu$ was computed directly from the
abundance ratios $\mathrm{N_H/N_{He}}$, and $\mathrm{N_Z/N_{He}}$ of
\citet{Nugis2000}.  This same value was used for the other WN8h star
considered, WR\,116, and in both cases $\gamma_e=1.1$ was assumed from
analogy with other late WN's.

\subsection{Total Neon Abundance}
\label{sec:total-neon-abundance}

Estimating the \emph{total} neon abundance from the spectra is
difficult, especially since the \neIII\ line at 15.56\um\ is outside of
the N-Band atmospheric window, and thus cannot be used to probe the gas
in a higher ionization state.  Theoretical neon ionization structure
predictions could in principle be used to infer total neon abundances.
Neon is now routinely included in the blanketed, non-LTE WR atmosphere
code CMFGEN \citep{Hillier1998}, and hence its ionization structure can
be modeled.  It is an impurity species, and generally only has a small
effect on the spectral energy distribution; it can, however, help drive
the wind, especially in the outermost regions.  Although a wind model
specific to the very late WC9 star WR\,121 was not available, by analogy
to other models for low ionization WR stars (e.g. WN10), \nep\ is likely
to be the dominant ion in the outer wind (J.  Hillier, 2004, private
communication).

Some handle on the most likely ionization state can also be had by
noting that the ionization potential of S$^{3+}$ is 35 eV, vs.  \nep:
21.6 eV and \nepp: 41.0 eV.  Since their critical densities are
reasonably close at these temperatures ($n_e\sim 10^5$), it is expected
that the detection of \sIV\ at 10.5\um\ is an excellent predictor of
\neIII.  Indeed, in the hot WN8 star WR\,147, \citet{Morris2000} found
\neIII\ and \sIV, but no \neII, and non-LTE model predictions of the
outer winds of late WC stars show Ne$^+$ as the dominant ion species
\citep{Willis1997}.  While the \emph{presence} of \sIV\ clearly does not
imply a dearth of Ne$^+$ \citepalias[cf.
WR\,105,][]{2001AJ....121.2115S}, its \emph{absence} places a stronger
limit on the possible existence of Ne$^{++}$.\footnote{Since sulfur
  abundance is not enhanced by nucleosynthesis and neon is, this
  argument is weaker if the \nes\ abundance ratio were so large that
  \sppp, while present, remained undetected, despite the strong measured
  \neII\ emission.}  By this argument Ne$^+$ accounts for nearly all the
neon in those stars with no \sIV\ detection.  Additional support for
this conculsion as it pertains to WR\,121 is provided by another WC9
star recently observed with the Spitzer's IRS spectrograph, which offers
similar resolution as \score\ but covers 10--40\um.  The IRS spectrum
shows strong \neII\ without any detectable 15.56\um\ \neIII.  This and
other early IRS WR results will be presented in a forthcoming paper.

\subsection{Cosmic Neon Abundance}
\label{sec:cosm-neon-abund}

The ``cosmic'' abundance of neon measured from solar coronal lines has
been revised many times over the past 30 years.  Recommended values for
the fractional abundance by number, $\epsilon_{Ne}=10^{12}\times
\mathrm{N_{Ne}/N_H}$ have ranged from $\lene=7.64$ to $8.04$
(converted from the total mass abundance of, e.g., \citet{Cameron1973}
and the references in \citet{Maeder1983}).  More recent values include
those suggested by \citet{Grevesse1998} ($\log(\epsilon_{Ne})=8.08$),
and the value found from the updated oxygen abundances of
\citet{Asplund2004} ($\log(\epsilon_{Ne})=7.84$).

For evolved WC stars, the cosmic abundance is of little interest, since
the bulk of the neon atoms entrained in the wind material were created
directly from helium burning in the core.  Final neon abundances with
respect to helium can then be compared directly to model predictions,
with little sensitivity on the initial amount of neon.

The less evolved WN stars do not produce any neon directly, and are
therefore expected to exhibit \nehe\ matching cosmic abundances.
However, all WR stars produce, and potentially consume, helium, so that
the ``cosmic'' value of \nehe\ actually varies, depending on what
assumptions are used to correct the helium abundance for its enhancement
and/or depletion via nuclear processing.  For example,
\citet{Morris2000} and \citet{Morris2004} derive a corrected \nehe\ 
abundance by assuming complete conversion from $4H\rightarrow He$ for
the WN targets considered:

\begin{equation}
  \label{eq:15}
  \left(\frac{\mathrm{Ne}}{\mathrm{He}}\right)_{corr}=
  \mathrm{\frac{N_{Ne}}{N_{He}+N_H/4}=\frac{Ne/H}{He/H+1/4}}
\end{equation}

\noindent Using the latest \citeauthor{Asplund2004} abundances, this yields a
cosmic abundance $\nehe=2\times10^{-4}$ in the severely H-depleted
environment of WN winds.  \citet{Barlow1988} perform a similar
adjustment for the WC8 star $\gamma^2$ Velorum, assuming C/He=0.2 to
correct \nehe\ for processing of all hydrogen to helium, and subsequent
conversion of helium to further elements.

Since any unprocessed hydrogen in the core or wind material (common in
late WN stars) will invalidate the complete conversion assumption of
Eq.~\ref{eq:15}, we adopt a range of cosmic \nehe\ using the updated
solar abundances of \citet{Asplund2004}, from $\nehe=2.0\times10^{-4}$
(complete H processing) to $3.7\times10^{-4}$ (no H processing).

\section{RESULTS}
\label{sec:results}

\begin{deluxetable*}{llccccccc}
\tablecolumns{9}
\tablewidth{0pc} 
\tablecaption{Neon and Sulfur Abundances.\label{tab:abund}}
\tablehead{
\colhead{Object} &
\colhead{Type} &
\colhead{$D$} &
\colhead{$\gamma_e$} &
\colhead{$\mu$} &
\colhead{$v_\infty$} &
\colhead{$\log(\dot{M})$} &
\colhead{Line Flux} &
\colhead{$\mathrm{\frac{X_i}{He}}$}\\
\colhead{}&
\colhead{}&
\colhead{(kpc)}&
\colhead{}&
\colhead{}&
\colhead{(km/s)}&
\colhead{($\mathrm{M_{\odot}/yr}$)}&
\colhead{($10^{-19}\:\mathrm{W/cm^2}$)}&
\colhead{($\times 10^{-4}$)}}

\startdata
\cutinhead{Ne$^+$}
WR\,105 & WN9h & 1.58 & 1.0 & 2.6 & 1200 & -4.55      & 1.5 & 8.8\\
WR\,116 & WN8h & 2.48 & 1.1 & 2.0 & 800  & -4.6       & 1.6 & 5.3\\
WR\,121 & WC9d & 1.83 & 1.1 & 4.7 & 1100 & -4.9\tablenotemark{\dag} & 2.1 & 41\\
WR\,124 & WN8h & 3.36 & 1.1 & 2.0 & 710  & -4.61      & 0.5 & 4.2\\
\cutinhead{S$^{3+}$}                                 
WR\,105 & WN9h & 1.58 & 1.0 & 2.6 & 1200 & -4.55      & 1.2\tablenotemark{\ddag} & 0.5
\enddata

\tablenotetext{\dag}{A radio limit of $\log(\dot{M})<-4.8$ is given by
  \citet{Abbott1986}, and we adopt the plausible value -4.9 based on other
  measured WC9 rates (e.g. WR\,80).}
\tablenotetext{\ddag}{Including 15\%
  correction for helium contamination of the line.}
\end{deluxetable*}

A summary of the inputs and results of the neon and sulfur abundance
calculations for the four program sources are given in
Table~\ref{tab:abund}.  Immediately apparent is the overabundance of
Ne$^+$ in WR\,121, the late WC star.  The three WN stars for which
Ne$^+$ abundance was measured are not expected to display abundance
enhancements, since the byproducts of CNO processing should dominate.
All are reasonably close to the cosmic value for \nehe, ranging from
1.1--4.4$\times$ cosmic (the latter corresponding to the fully-processed
cosmic value), with WR\,105 noticeably higher than the others.

Presuming all neon is accounted for in Ne$^+$, the total neon abundance
can still be increased by introducing the clumping fill factor $\delta$.
If $\delta=0.1$, the abundances are increased by $\sim$1.8, presuming
all other background abundances are unaffected.  If all abundances
(including helium) scale the same with clumping factor, the abundance
ratios should technically be insensitive to it, but differing emission
regimes complicate this argument (see
\S\,\ref{sec:neonsulfur-abundance}).

Another interesting constraint is available by considering the mass loss
rate dependence.  If for, e.g., the nitrogen star WR\,124, the
Optical+UV line analysis based rate of \citet{Hamann1998b}
($\dot{M}=10^{-3.8}\msun\,\mathrm{yr}^{-1}$) had been adopted with all
other inputs unchanged, the derived abundance would drop to
$\nephe=2\times 10^{-5}$, or only $\sim$5\% of the cosmic neon value.
Since the reactions which convert neon to magnesium occur so rarely, and
at core temperature attained only at latest stages of WR evolution and
for the most massive WR cores, this casts doubt on such a high mass loss
measurement; once predicted wind abundances are confirmed, neon and
other products of nuclear processing can be used to constrain mass loss
rates in the absence of other measures.

The WC star WR\,121's neon abundance $\nephe=4.1\times 10^{-3}$, is in
quite good agreement with the long-standing prediction of $\nehe=6.6
\times 10^{-3}$.  Though any significant abundance of Ne$^{++}$, which
we could not detect, would serve to increase \nehe, doubly ionized neon
is not expected to be abundant in this source, due to the total absence
of \sIV\ emission.  A realistic clumping factor of $\delta=0.15$ would
imply an abundance exactly equal to the predicted value.  

Sulfur, though produced in the advanced oxygen burning leading up to
supernovae, is not significantly enhanced by normal stellar
nucleosynthesis, and exhibits no abundance increase.


\section{DISCUSSION}
\label{sec:discussion}

\subsection{Neon Detection Frequency}
\label{sec:neon-entire-wr}


Of the 29 WR stars cataloged by \citetalias{2001AJ....121.2115S}, only four
exhibited measurable \neII\ emission.  The absence of neon in the
remaining 25 is likely due to two factors: higher typical ionization of
the wind in the neon-emitting regime, and bright continuum from heated
dust diluting any spectral lines which otherwise would be present.

Among the ISO WR stars with detected neon emission, only $\gamma^2$ Vel
(WC8) showed any non-nebular \neII.  The remaining WC5--WC8 and WN8 ISO
spectra exhibited only \neIII\ 15.55\um, which is inaccessible in the
ground-based \score\ data.  That the low \neII\ detection frequency is a
result of higher average ionization is supported by the late subtypes of
the four program stars discussed here; for both WN and WC, the subtype
sequence is primarily one of ionization, with early types exhibiting
lines of increasingly higher ionization species.  This same
earlier-type, increasing ionization sequence is also found in the trends
of helium line emission strengths in the MIR spectra of
\citetalias{2001AJ....121.2115S}.

The majority of WC9 stars are dust emitters, with infrared excesses
10$\times$ or more the normal free-free excess seen in non-dust
producing WR stars \citep{Williams1987}.  The bright infrared continuum
of heated carbon dust is usually unaccompanied by line emission in the
thermal IR; presumably the lines suffer so much continuum dilution they
are unmeasurable.  Among the 6 WC9d late-type WC's observed by
\citetalias{2001AJ....121.2115S}, only WR\,121 showed any line emission,
and indeed it is the only WC9 for which any MIR line emission has ever
been observed (the two neon-emission WC10 stars reported by
\citet{Aitken1980} based on early low-resolution ground-based
spectroscopy were later revealed be planetary and proto-planetary
nebulae).

A notable star in the original MIR sample of
\citeauthor{2001AJ....121.2115S} without \neII\ present is WR145, a
WN/WC type.  The rare WN/WC transition stars, eight of which are
cataloged by \citet{vanderHucht2001}, exhibit spectral signatures
intermediate between the WN and WC classes, and are hypothesized to
represent a transitional stage between CNO-dominated hydrogen-free WN
stars, and $\alpha$-burning dominated WC stars \citep*{Crowther1995a}.
A confirmation of enhanced neon abundance in one of the transition
objects would support this interpretation, but unfortunately any neon in
WR145 must be present as \nepp, as expected from the strong \sIV\ 
emission the star exhibits.

\subsection{Neon/Sulfur Abundance}
\label{sec:neonsulfur-abundance}

As pointed out by \citet{Dessart2000}, the use of neon abundances as
derived in the previous section to constrain true core-evolution nuclear
processes is complicated by the different emitting regimes of the
elements in question.  Because of their comparatively small radiative
$A$ coefficients, the fine structure lines originate further out in the
wind than the helium emission, at densities near
$10^5\,\mathrm{cm}^{-3}$.  A dependence on precise mass loss rates and
(less problematic) terminal velocities also reduces the accuracy of
direct abundance measurements, especially given the tendency of clumping
to modify the measured $\dot{M}$ by factors of 2--5.  A method which
overcomes these difficulties is available in the \nes\ abundance.  Since
both \sIV\ and \neII\ originate in roughly the same region of the wind,
the abundance ratio derived from them should be independent of details
of the wind structure, and remove any attendant systematic errors in
estimating the mass loss rate, distance, and bulk parameters of the
star.

The \heI\ and \heII\ lines blended with \sIV\ 10.5\um\ are known by
wind ionization models to contribute only very weakly to this line in
$\gamma^2$ Velorum \citep[$<$15\%,][]{DeMarco2000}.  Due to the blend of
neutral and ionized helium and hydrogen lines at this location, this
contamination fraction is expected to remain nearly constant with WR
subtype.  We have therefore reduced the measured \sIV\ input flux in
Table~\ref{tab:abund} by this amount.

The \sppphe\ abundance found for the WN9 star WR\,105 is in good
agreement with the cosmic value of $\she=7.5\times 10^{-5}$, and if
S$^{++}$ constitutes about one-third of the sulfur, as in WR\,146
\citep{Willis1997}, the total sulfur abundance comes quite close to
cosmic.  The neon to sulfur ratio we find is $\nepsppp=15.7$, somewhat
larger than the cosmic value of the full abundance ratio $\nes=7$.  If,
as expected, sulfur is less completely accounted for by S$^{3+}$ than
neon is by Ne$^+$, the total sulfur abundance would be increase by a
larger factor than would neon (around a factor of 2), and the final
implied \nes\ can approach the expected cosmic value, as expected in
this less-evolved WN star.

\section{Conclusions}
We have computed neon abundances for 3 WN and 1 WC star from
ground-based spectra of the \neII\ 12.81\um\ emission line.  WR\,121,
the WC star, shows elevated \nep\ abundance, with an estimated total
neon 11.1$\times$ the cosmic value, close to the expected 17.8$\times$
increase predicted by WR core evolution models.  A significant
population of emitting neon ions in the \nepp\ ionization level, or
realistic clumping fill factors $\delta$ would each revise this value
upwards by up to a factor of 2, but distance and mass-loss rate
uncertainties also contribute.  Though elevated neon abundances have
been found in several other WC stars, this is the least-evolved star for
which such enhancement has been demonstrated.  The WN stars were found
to have abundances close to cosmic, consistent with no nuclear neon
enhancement.  For the single WN star for which neon and sulfur were both
observed, the \nes\ ratio, which is insensitive to uncertainties in the
star's bulk parameters, was found to be consistent with cosmic values,
despite larger uncertainties in the total sulfur abundance.

\acknowledgments

We are grateful for the assistance and comments provided by Andre
Maeder, Luc Dessart, Pat Morris, and Goetz Graefener.  We thank John
Hillier for his helpful discussion of neon ionization structure.  We
also thank the staff of Palomar Observatory for their dedicated support.
JDTS acknowledges support from NASA through JPL contract 1224769.

\bibliographystyle{apj}
\bibliography{general,myref}
\end{document}